\documentclass[twocolumn,showpacs,preprintnumbers,superscriptaddress,
amsmath,amssymb]{revtex4}

\usepackage{epsfig}
\usepackage{graphicx}
\usepackage{dcolumn}
\usepackage{bm}
\usepackage{times}
\usepackage{xcolor}

\usepackage{cases}

\begin{document}

\title{Preferential imitation of vaccinating behavior can invalidate the targeted subsidy on complex networks}

\author{Hai-Feng Zhang}

\affiliation{School of Mathematical Science, Anhui University, Hefei
230039, P. R. China}
\affiliation{Key Laboratory of Computer Network and Information Integration (Southeast University), Ministry of Education}
\affiliation{Department of Communication Engineering, North
University of China, Taiyuan, Shan'xi 030051, P. R. China}

\author{Pan-Pan Shu}
\author{Ming Tang}
\email{tangminghuang521@hotmail.com}
\affiliation{ Web Sciences Center, University of Electronic Science
and Technology of China, Chengdu 611731, P. R. China}

\author{Michael Small}\email{michael.small@uwa.edu.au}
\affiliation{School of Mathematics and Statistics, The University of Western Australia, Crawley, Western Australia 6009,
Australia}

\date{\today}

\begin{abstract}
We consider the effect of inducement to vaccinate during the spread of an infectious disease on complex networks. Suppose that public resources are finite and that only a small proportion of individuals can be vaccinated freely (complete subsidy), for the remainder of the population vaccination is a voluntary behavior --- and each vaccinated individual carries a perceived cost. We ask whether the classical targeted subsidy strategy is definitely better than the random strategy: does targeting subsidy at individuals perceived to be with the greatest risk actually help? With these questions, we propose a model to investigate the \emph{interaction effects} of the subsidy policies and individuals responses when facing subsidy policies on the epidemic dynamics on complex networks. In the model, a small proportion of individuals are freely vaccinated according to either the targeted or random subsidy policy, the remainder choose to vaccinate (or not) based on voluntary principle and update their vaccination decision via an imitation rule. Our findings show that the targeted strategy is only advantageous when individuals prefer to imitate the subsidized individuals' strategy. Otherwise, the effect of the targeted policy is worse than the random immunization, since individuals preferentially select non-subsidized individuals as the imitation objects. More importantly, we find that under the targeted subsidy policy, increasing the proportion of subsidized individuals may increase the final epidemic size. We further define social cost as the sum of the costs of vaccination and infection, and study how each of the two policies affect the social cost. Our result shows that there exist some optimal intermediate regions leading to the minimal social cost.

\end{abstract}


\maketitle

\begin{quotation}
\end{quotation}

\section{Introduction} \label{sec:intro}
Vaccination is the most effective response to large-scale epidemic
outbreaks, such
as seasonal influenza and influenza like epidemics. Vaccination both limits the extent of the outbreak and reduces morbidity and mortality~\cite{reluga2006evolving,mbah2012impact}. However, electing to take vaccination brings certain side effects or inconveniences; while
protecting not only those who are vaccinated but also their
neighbors, and  leading to the so-called effect of
``herd-immunity''. This scenario naturally leads to the problem of
``free-riding'' commonly observed in public goods
studies~\cite{wu2011imperfect}.
Recently, by incorporating evolutionary game theory into traditional epidemiological models, many researchers have demonstrated that a voluntary vaccination policy without incentives may
be unable to eradicate a vaccine-preventable disease
~\cite{haifeng2013braess,reluga2011general,zhang2010hub,bauch2003group,bauch2004vaccination,
bauch2005imitation,vardavas2007can}. One possible way
of solving the social dilemma with respect to vaccination is for the
government or health organizations to offer some incentive
programs, e.g., subsidy and insurance policies~\cite{zhang2012impacts,wells2013policy}.

Since many diseases spread through human populations by
physical contact between infected individuals and susceptible individuals, the pattern of
these disease-causing contact forms a network. In the past decades, network-based epidemiological models have attracted myriad attention and many significant results have
been achieved~\cite{newman2002spread,moreno2002epidemic,BBPSV:2005,PhysRevLett.105.218701,yang2008selectivity,boccaletti2014structure,granell2013dynamical}. To control the epidemic spreading on complex networks, many well-studied immunization strategies have been proposed, such as targeted immunization~\cite{pastor2002immunization} and
acquaintance immunization~\cite{cohen2003efficient}. Compared to a random immunization strategy, these two immunization strategies have been proven to be more efficient in controlling the spread of epidemic diseases on complex networks. However, the results were obtained by the two implicit assumptions: 1) there are enough public resources to vaccinate many individuals; 2) the behavioral responses of the remaining individuals are not considered. In reality, public resource limitations mean that government agencies can usually only freely vaccinate a small proportion of people. The remaining population decide whether or not to take vaccination according to a voluntary principle --- and subject to an associated vaccination cost. In this case, the vaccination decision of the remaining people is dependent on the risk of infection, prevalence of infection, the cost of vaccination, public opinion and so on~\cite{funk2009spread}. In particular, their vaccination decisions are sensitively influenced by the proportion of subsidized individuals as well as which kinds of individuals are subsidized. For example, if some hub nodes are freely immunized, the nodes with small degrees have small probability of being infected due to the herd immunity effect from these hub nodes, as a result, many nodes are unwilling to take vaccination, leading to an increase in the final epidemic size. Therefore, we need to reconsider whether the classical targeted subsidy policy (i.e.,  targeted immunization under which certain individuals with the highest  degree are freely immunized) is still better than the random case~\cite{wells2013policy,rat2010modelling}.

In this paper, a vaccination decision model with a preferential selection mechanism
based on game theory is established to investigate the interplay of the subsidy policy, the voluntary vaccination decision and the spreading dynamics on networks. We then compare the effects of the targeted subsidy policy with the random policy on the vaccination and epidemic dynamics. We find that the preferential selection mechanism plays a vital role on whether the targeted subsidy policy yields ideal results. We show that the advantage of the targeted policy is greatly increased when the non-subsidized individuals prefer to select the subsidized individuals as imitation objects. However, the control effect of the targeted policy is even worse than that of the random policy when non-subsidized individuals randomly or preferentially select other non-subsidized nodes as their imitation objects. More importantly, we find that, for the targeted policy, increasing the subsidy proportion may increase \emph{rather than} reduce the final epidemic size when non-subsidized individuals are preferentially selected, it is because that a small subsidy proportion may greatly reduce the vaccination willingness of the non-subsidized individuals. In addition, by defining the social cost as the sum of the costs of vaccination and infection, we find that the social cost non-monotonically depends on the subsidized proportion $p$ for both subsidy policies. Our findings indicate that the effects of well-designed subsidy policies may be undermined or counter-productive if the policies are implemented without considering
the intricate human behavioral responses, such as neglecting the non-subsidized individuals' responses on the subsidy policies.

In Sec.~\ref{sec:model}, we describe our model of epidemic spreading
on complex networks by integrating game theory. In Sec.~\ref{sec:theory},
we present our main results concerning the effects of the two subsidy policies on the vaccination coverage, the final epidemic size and the social cost. Meanwhile, an example of influenza case is given in Sec.~\ref{sec:apendix} to demonstrate the applicability of these results. In Sec.~\ref{sec:conclusion},
we present conclusion and discussions.

\section{Descriptions of the model} \label{sec:model}

Taking into account seasonal flu-like diseases
and the limited effectiveness of vaccines and following previous studies
~\cite{fu2011imitation,haifeng2013braess,zhang2014effects}, we model the vaccination dynamics and the epidemic as a two-stage process: the first stage is the vaccination decision stage, which is implemented before each epidemic season; the second stage is the epidemic season stage, where non-vaccinated individuals may be infected by diseases described by the standard
susceptible-infected-removed (SIR)
model. Specifically, at the vaccination stage, except for a proportion $p$ of individuals who are freely vaccinated (according to either the targeted strategy or the random strategy), each individual needs to decide whether or not to take vaccination. Choosing to take vaccination  will incur a cost $C_V$ [the cost of vaccination can account for the monetary cost of the vaccine, the
perceived vaccine risks, side effects, long-term healthy impacts,
and so forth] regardless of the non-subsidized or subsidized individuals (the vaccination cost of the subsidized individuals is $C_V$ though they need not pay the cost themselves \footnote{Of course, this assumes that the non-monetary as well as monetary cost of vaccination can be reduced to zero for the subsidised vaccinations.}). We also assume that the vaccine can perfectly protect vaccinated individuals from infection in the following epidemic season. The second stage is the epidemic season stage, during which each unvaccinated individual
has a probability of being infected during the season.
In this stage the epidemic strain infects an initial number $I_0$ of
individuals, and the disease spreads according to SIR dynamics with daily
transmission rate $\lambda$ and recovery rate $\mu$. The epidemic
continues until there are no more newly infected individuals. An
individual who gets infected during the epidemic season pays a
cost $C_I$. Without loss of generality, we set $C_I =1$ and let $c=C_V/C_I$
describe the relative cost of vaccination, whose value is
restricted in the region of $[0,1]$~\cite{fu2011imitation,liu2012impact}. Those individuals who are neither
vaccinated nor infected pay no cost, so they are the free riders.

Let $P_i(t)$ be the current payoff of individual
$i$ at season $t$. According to the above descriptions, $i$ takes one of the three cost values as follows
\begin{numcases}{P_i(t)=}\label{1}
\nonumber -c,~~~~~~~~~~~~~ \mbox{vaccination};\\
-1 ,~~~~~~~~~~~~\mbox{infected};\\
\nonumber 0,~~~~~~~~~~~~~~~\mbox {free-rider}.
\end{numcases}

After the last epidemic season, individuals can update their strategies again in the following vaccination decision stage based on their previous experiences.
Considering that imitation phenomena commonly exist in the real world~\cite{tennie2006push,heyes1996social}, many studies have investigated the impact of an imitation rule from the game theoretical perspective as well as in the vaccination decision-making epidemiological models~\cite{fu2011imitation,mbah2012impact}. Hence, we also assume individuals update their strategies based on an imitation rule. Specifically,
if an individual $i$ updates her vaccination strategy, she \emph{randomly} chooses an immediate neighbor $j$ with which to compare their expected payoff, and then adopts $j$'s strategy with a probability dependent on
their payoff difference~\cite{fu2011imitation,szabo1998evolutionary}:
\begin{equation}\label{2}
W(s_i\leftarrow s_j)=\frac{1}{1+\exp[-\beta(P_j(t)-P_i(t))]},
\end{equation}
where $s_i = 1 \ \mbox{or} \ 0$ denotes the vaccination choice for individual
$i$: either vaccinated or not,
where $\beta$ describes the rationality of the individuals.
Unless otherwise stated, we let $\beta=10$ in this paper, implying that
better-performing individuals are readily imitated, but it is still possible to adopt the behavior of an individual performing
worse.

When the subsidy policies from the governments or health organizations are included, it is difficult to be definitive over whether individuals choose to imitate subsidized or non-subsidized individuals. For example, some individuals may prefer to imitate the subsidized individuals' strategies since these subsidized individuals are chosen by authoritative organizations. Conversely, some individuals are willing to imitate  non-subsidized individuals' strategies since their strategies come from their self-choices but not from the external environments. In view of this, we introduce a selection parameter $\alpha$ to differentiate the preferential selection rule. In detail, we first give a different weight value $w=1$ ($w=5$) for non-subsidized (subsidized) individuals\footnote{Here we use 1 and 5 to distinguish the non-subsidized and subsidized individuals, respectively. Obviously, the other two different positive values have the same effects.}, then the probability of node $i$ choosing neighbor $j$ is given as:

\begin{equation}\label{2}
f(i\leftarrow j)=\frac{e^{\alpha w_{j}}}{\sum_{k\in\Lambda_i}e^{\alpha w_k}},
\end{equation}
here $\Lambda_i$ denotes the neighborhood of node $i$ and $w_i$ is the weight of node $i$. For $\alpha<0$, non-subsidized neighbors are more likely to be selected as the potential strategy objects. On the contrary, for $\alpha>0$, individuals tend to select subsidized neighbors as potential imitation objects. Neighbors are randomly selected as $\alpha=0$, which returns to the original model. Our main aim is to explore how the value of $\alpha$ affects the vaccination coverage, the final epidemic size and the social cost in regard with the two different subsidy policies.

\section{main results} \label{sec:theory}
In our simulations, for targeted subsidy policy, a proportion $p$ of individuals with the highest degrees are freely vaccinated, for the random subsidy policy, a proportion $p$ of individuals of individuals is randomly selected and freely vaccinated.
Then equal fractions of
vaccinated and non-vaccinated individuals are randomly distributed
among a proportion $1-p$  of individuals to start the iterative process. Meanwhile, the number of initially infected
individuals in each epidemic season is $I_0=5$.  The
long run equilibrium results are obtained by averaging the last 1000 iterations of a total of 3000 iterations in
100 independent realizations. Our results are implemented on networks generated from the
standard configuration model~\cite{newman2001random} with degree
distribution $P(k)\sim k^{-3}$. The size of the network studied
is $N =2000$, the minimal and maximal degrees are $k_{min}=3$ and
$k_{max}=\sqrt{N}$, respectively. Finally, in Sec. \ref{sec:apendix} we also study a practical case by using the email network to validate the generality of the results and demonstrate the application of this approach.

Before comparing the two subsidy policies' effects on the vaccination coverage ($V$) and the final epidemic size ($R$), we first investigate how the pure immunization strategy (we also say subsidy policy in the following context, only $p$ proportion of individuals is immunized, the remaining individuals do nothing) and the pure vaccination decision dynamics (without the subsidy policy, \emph{i.e.}, $p=0$) affect the epidemic dynamics, respectively. The targeted immunization strategy (labelled TAR in the following figures) and the random immunization strategy (labelled RAN in the following figures) are compared in Fig.~\ref{fig1}(a). As provided in previous work, Fig.~\ref{fig1}(a) shows that the targeted immunization strategy can better control the outbreaks of epidemics when the resources are sufficient, and can be eliminated completely once $p\approx0.18$. Furthermore, the vaccination coverage $V$ and the final epidemic size $R$ versus the cost of vaccination $c$ are given in Fig.~\ref{fig1}(b), one can find that the value of $V$ decreases with the value of $c$, leading to a corresponding increase of $R$.
\begin{figure}
\begin{center}
\includegraphics[width=3.5in]{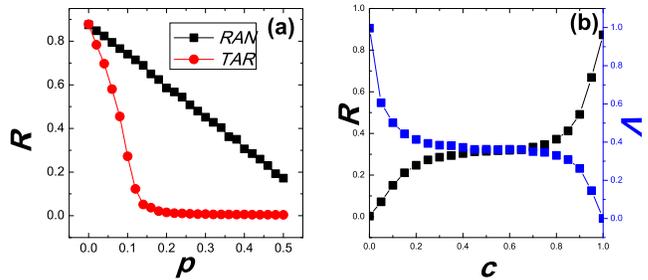}
\caption{(Color online) (a) For the pure immunization strategy, comparison of the effects the targeted immunization (or subsidy) strategy and the random immunization strategy on the final epidemic size $R$; (b) the final epidemic size $R$ and the vaccination coverage $V$ as functions of the cost of vaccination $c$ when the subsidy proportion $p=0$. In this and the following figures, unless otherwise specified, we let $\lambda=0.2$ and $\mu=0.25$ to calibrate the epidemic size near $0.9$ when there are no protective measures.}
\label{fig1}
\end{center}
\end{figure}

\begin{figure}
\begin{center}
\includegraphics[width=3.5in]{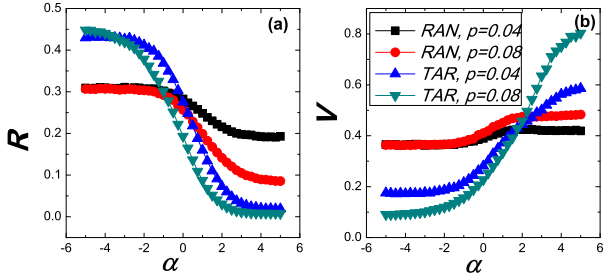}
\caption{(Color online) Effects of $\alpha$ on the control effects of the random subsidy policy and the targeted subsidy policy for different values of $p$:  (a) The final epidemic size $R$ versus the select parameter $\alpha$ for different subsidy policies; (b) the vaccination coverage $V$ versus the select parameter $\alpha$ for different subsidy policies. Here the cost of vaccination $c=0.5$.}
\label{fig2}
\end{center}
\end{figure}

In the following, we want to know what new phenomena can be produced when the subsidy policies and the vaccination decisions from an individual's voluntary behaviors are integrated. Fig.~\ref{fig2} gives the final epidemic size (see Fig.~\ref{fig2}(a)) and the vaccination coverage (including the subsidized individuals and the vaccinated individuals decided by themselves, see Fig.~\ref{fig2}(b)) as functions of the parameter $\alpha$ for different subsidy policies as well as the different values of $p$. As shown in Fig.~\ref{fig2}(b), for $\alpha>0$, the vaccination coverage is enhanced and the larger value of $p$ gives rise to the greater enhancement of $V$ for the both subsidy policies. However, one can see that the vaccination coverage for the targeted policy is enhanced more significantly, leading to almost the elimination of the epidemic for the targeted policy (see Fig.~\ref{fig2}(a)). On the contrary, for $\alpha\leq0$, a non-trivial phenomenon appears: the vaccination coverage for the targeted policy is much lower than the random policy, meanwhile, for the targeted policy, larger value of $p$ yields lower level of vaccination coverage. The non-trivial phenomenon becomes more remarkable as the value of $\alpha$ is further decreased. Correspondingly, the control effect of the targeted policy becomes worse than the random case. This result suggests that the feel good subsidy policy (targeted subsidy policy) may produce the opposite effect when the individuals' responses on subsidy policies are neglected.

\begin{figure}
\begin{center}
\includegraphics[width=3.5in]{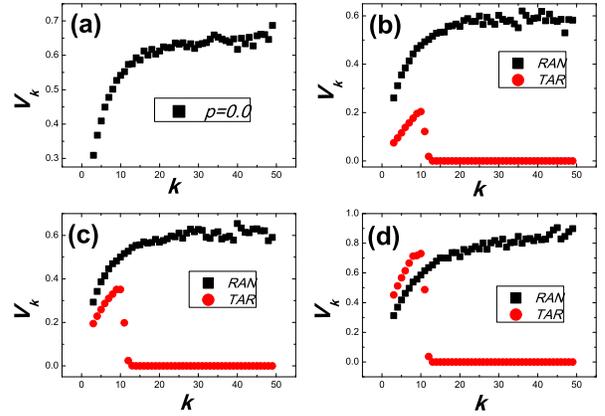}
\caption{(Color online) The voluntary vaccination probability $V_k$ as a function of degree $k$ for the random subsidy policy and the targeted subsidy policy. (a) the subsidy policy is not considered, i.e., $p=0$. (b) $\alpha=-3$ and $p=0.05$; (c) $\alpha=0$ and $p=0.05$; (d) $\alpha=3$ and $p=0.05$.  Here the cost of vaccination $c=0.5$.}
\label{fig3}
\end{center}
\end{figure}

To systematically analyze the above phenomena, we first define the voluntary vaccination probability $V_k$ (excluding the subsidized individuals) for the nodes with the same degree $k$ as: $V_k=N^{v}_{k}/N_k$, where $N^v_k$ and $N_k$ are the \emph{number} of voluntarily vaccinated nodes with degree $k$ and the number of nodes with degree $k$, respectively.  In Fig.~\ref{fig3}(a), we plot $V_k$ as a function of $k$ for the case of $p=0$, \emph{i.e.}, the subsidy policy is omitted. As illustrated in Fig.~\ref{fig3}(a), the probability $V_k$ increases with the degree $k$ until to a upper bound ($\approx0.65$), that is to say, the nodes with larger degrees (`large nodes' or hub nodes) are more likely to take vaccination than the `small nodes' (the nodes with small degrees) even when there is no any subsidy policy. The reason comes from that the `large nodes' have higher risk of infection~\cite{zhang2010hub}.

We then consider the cases of $\alpha=-3$, $0$ and $3$ in Fig.~\ref{fig3}(b), (c) and (d) with $p=0.05$. One can observe that, when $\alpha<0$ (Fig.~\ref{fig3}(b)) or $\alpha=0$ (Fig.~\ref{fig3}(c)), the value of $V_k$ for the random subsidy policy is larger than the targeted case. Note that, for the targeted case, $V_{k\geq13}=0$ since the nodes with degree $k\geq13$ have been freely vaccinated. As analyzed in Fig.~\ref{fig3}(a), the voluntary vaccination probability for hub nodes is high even without the subsidy policy, in this situation, the governments or health organizations still use the limited public resource to subsidize these nodes rather than the nodes who are not willing to take vaccination voluntarily, meaning that the public resources are somewhat wasted and cannot as fully stimulate the vaccination enthusiasm of `small nodes'. Meanwhile, for the targeted policy, the non-subsidized nodes are `small nodes', and they are not willing to take voluntarily vaccination owing to the low risk of infection. In this case, the non-vaccinated neighbor nodes have higher probability of being selected if $\alpha<0$, which leads to the low vaccination coverage for the targeted policy, \emph{i.e.}, the lower value of $V_k$. However, for the random subsidy policy, each node has the same probability to be freely vaccinated, in this case, the hub nodes with the higher voluntarily vaccination probability [see Fig.~\ref{fig3}(a)] are more likely to be selected by others owing to their large connectivity even when $\alpha\leq0$, so their vaccination decision is more likely to be imitated by others. Combining the above explanations, we can understand why the vaccination coverage for the targeted policy is much lower than the random case as $\alpha\leq0$. When $\alpha>0$ (Fig.~\ref{fig3}(d)), firstly, subsidized individuals are likely to be selected as the imitation objects, secondly, for the targeted subsidy policy, the subsidized nodes are the hub nodes who have more connectivity and can be selected by others with larger probability. As a result, the $V_k$ of the targeted policy is larger than the random case.

\begin{figure}
\begin{center}
\includegraphics[width=3.5in]{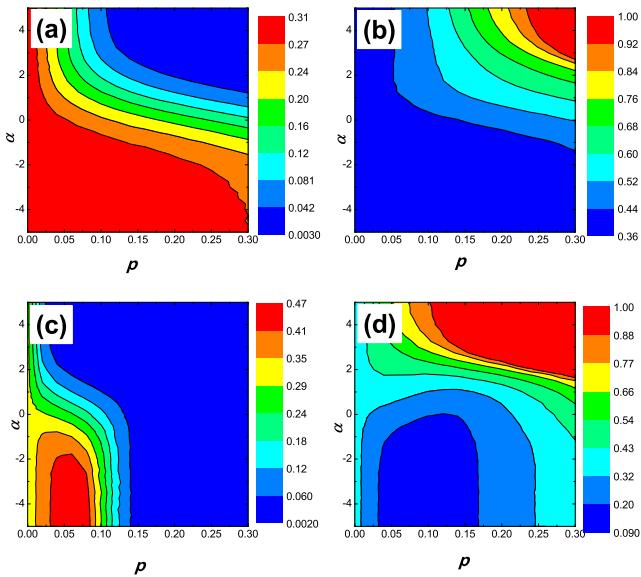}
\caption{(Color online) Comparison between the random subsidy policy and the targeted subsidy policy for the final epidemic size $R$ and the vaccination coverage $V$ in the full phase diagram ($\alpha,p$).(a) The final epidemic size $R$ versus $\alpha$ and $p$ for random policy; (b) the vaccination coverage $V$ versus $\alpha$ and $p$ for random policy; (c) the final epidemic size $R$ versus $\alpha$ and $p$ for targeted policy; and, (d) the vaccination coverage $V$ versus $\alpha$ and $p$ for targeted policy. Here the cost of vaccination $c=0.5$.}
\label{fig4}
\end{center}
\end{figure}

Furthermore, Fig.~\ref{fig4} plots the contour of $R$ (left column of Fig.~\ref{fig4}) and the contour of $V$ (right column of Fig.~\ref{fig4}) versus the parameters $\alpha$ and  $p$ for the two subsidy policies. For the random subsidy policy, one can find that the vaccination coverage $V$ increases with $p$ and $\alpha$ (see Fig.~\ref{fig4}(b)), leading to the reduction of $R$ when $p$ or $\alpha$ is increased (see Fig.~\ref{fig4}(a)). For the targeted case, a very interesting result can be observed: for $\alpha<0$, the vaccination coverage first decreases with the parameter $p$ and to a minimal level as $p\in(0.03, 0.17)$, and then the vaccination coverage increases again with further increase in the value of $p$ (see Fig.~\ref{fig4}(d)).  Fig.~\ref{fig2}(b) and Fig.~\ref{fig3}(b) have shown the targeted subsidy policy cannot improve but will inhibit the vaccination coverage as $\alpha\leq0$, which can explain why the vaccination coverage is reduced at first. With the further increase of $p$, there are more vaccinated individuals even when the number of the voluntarily vaccinated individuals is negligible, so the vaccination coverage is increased again. The existence of the minimal vaccination coverage leads to the maximal final epidemic size as $p\in(0.03,~0.08)$ and $\alpha<-2$(see Fig.~\ref{fig4}(c)), which means that subsidy policy ($p>0$) does not reduce the final epidemic size but makes the final epidemic size larger than the case of $p=0$. Moreover, one can see that the maximal final epidemic size region (see Fig.~\ref{fig4}(c)) is much smaller than the minimal vaccination coverage region (see Fig.~\ref{fig4}(c)). The reason can be deduced from Fig.~\ref{fig1}(a), which indicates that the epidemic has been controlled to a rather low level as $p=0.1$.

From the perspective of group interest, the purpose of introducing subsidy policies is to minimize the total social cost at the
population level. We thus want to know how the social cost is affected by the two subsidy policies as well as the selection parameter $\alpha$. To this end, we first define the social cost as~\cite{dybiec2004controlling,zhang2012impacts}:
$SC=N_R\times1.0+N_V\times c$ with $N_R$ and $N_V$ be the \emph{number} of infected and vaccinated individuals.

\begin{figure*}
\begin{center}
\includegraphics[width=7in]{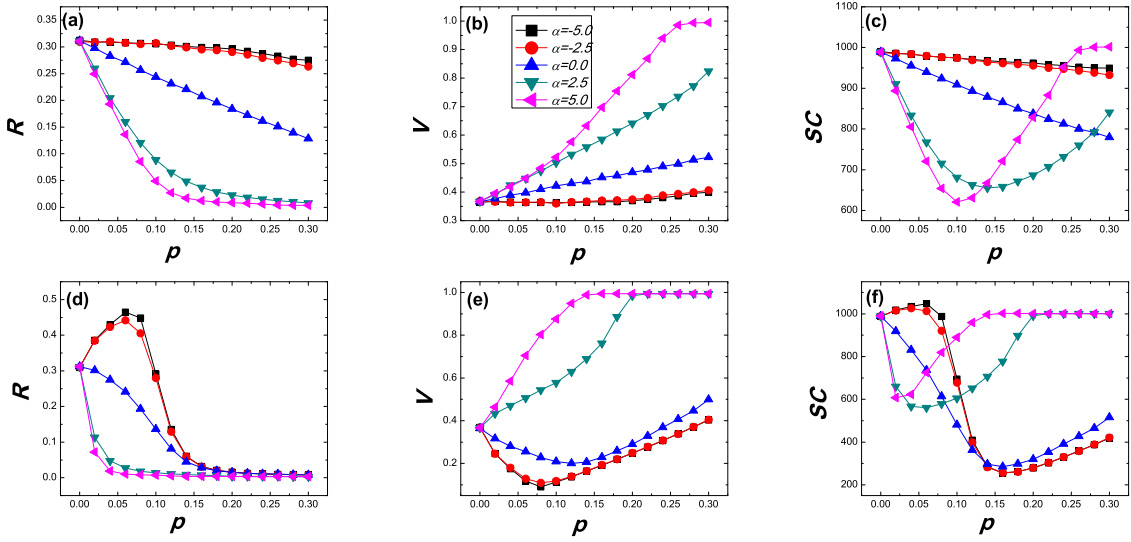}
\caption{(Color online) Given different values of $\alpha$, the final epidemic size $R$, the vaccination coverage $V$ and the social cost $SC$ respectively as a function of $p$ for the random policy and the targeted policy. (a) $R$ versus $p$ for the random policy; (b) $V$ versus $p$ for the random policy; (c) $SC$ versus $p$ for the random policy; (d) $R$ versus $p$ for the targeted policy; (e) $V$ versus $p$ for the targeted policy; (f) $SC$ versus $p$ for the targeted policy.  Here the cost of vaccination $c=0.5$.}
\label{fig5}
\end{center}
\end{figure*}

In Fig.~\ref{fig5}(c) we study the effect of the parameter $p$ on the value of $SC$ for the random subsidy policy. From Fig.~\ref{fig5}(c) we can find that the value of $SC$ decreases with $p$ as $\alpha\leq0$ (we here restrict $p\leq0.3$ given that the public resources are limited.). However, there exists an optimal value of $p$ leading to the minimal social cost when $\alpha>0$ (2.5 and 5.0). To explain such a phenomenon, we present the values of $R$ and $V$ as the functions of $p$ for different values of $\alpha$ in Fig.~\ref{fig5}(a) and Fig.~\ref{fig5}(b). One can see that the vaccination coverage $V$ is not increased remarkably as $\alpha<0$, which leads to the final epidemic size $R$ not being significantly reduced. Thus we need to further increase $p$ to reduce the social cost since the cost of infection is larger than the cost of vaccination. Yet, for $\alpha>0$, from Fig.~\ref{fig5}(b) and Fig.~\ref{fig5}(a) we can observe that the vaccination coverage is greatly improved and the $R\rightarrow0$ once $p>0.1$, which implies that the proportion of subsidy is sufficient to control the outbreak of epidemic, and further increase the value of $p$ just gives rise to the overload of subsidy.

The effect of subsidy proportion $p$ on the social cost $SC$ for the targeted policy is illustrated in Fig.~\ref{fig5}(f). Note that, compared with Fig.~\ref{fig5}(c), there also exists an optimal value of $p$ leading to the minimal social cost even for $\alpha<0$. As we know, unlike the random subsidy policy, Fig.~\ref{fig1}(a) shows that the epidemic is almost eliminated by the targeted subsidy policy as $p>0.15$ even without voluntarily vaccinated individuals. So further increase of $p$ definitely causes a rise in the social cost. For $\alpha>0$, the vaccination coverage increases with $p$ (see Fig.~\ref{fig5}(e)), leading to the rapid decrease of the final epidemic size, especially, when $p \approx0.05$, the epidemic is almost controlled (see Fig.~\ref{fig5}(d)). As a result, for $\alpha>0$, there exists an optimal value of $p$ around 0.05 causing the minimal social cost. Meanwhile, Fig.~\ref{fig5}(e) suggests that the optimal value of $p$ for $\alpha>0$ is smaller than the case of $\alpha\leq0$.

\begin{figure}
\begin{center}
\includegraphics[width=3.5in]{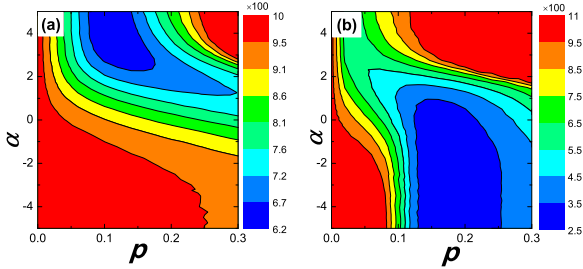}
\caption{(Color online) The contour of $SC$ as a function of ($\alpha, p$) for the random subsidy policy (a) and the targeted subsidy policy, respectively.  Here the cost of vaccination $c=0.5$.}
\label{fig6}
\end{center}
\end{figure}

The systematic effects of the parameters $p$ and $\alpha$ for the random policy and the targeted policy on the social cost are presented in Fig.~\ref{fig6}(a) and (b), respectively. As shown in Fig.~\ref{fig6}(a), the optimal phenomenon only evident when $\alpha>0$. However, for the targeted case, there always exists an optimal value of $p$ inducing the minimal social cost. Meanwhile, the optimal value of $p$ becomes larger as the value of $\alpha$ is gradually reduced.

\section{application: A practical example of influenza} \label{sec:apendix}

To demonstrate the universality of this phenomenon, we calibrate the parameters of the model based on the scenario of the 2009
H1N1 influenza epidemic, in which a reproduction number $R_0$ was
estimated as $R_0=1.6$~\cite{xia2013computational,yang2009transmissibility,poletti2011effect} and recovery rate was set to $\mu=1.0/3$ (\emph{i.e.}, the mean infectious period is 3 days)~\cite{fu2011imitation}. It is difficult to obtain the real contact networks on which the influenza spreads, so without loss of generality, we use an online social network---an email network as a proxy from which to study the spreading of influenza on social networks. The email network is composed of $N=1133$ nodes and 5451 links. In addition, the cluster coefficient, the assortative coefficient and the average shortest distance are respectively $C=0.11$, $r=0.078$ and $D=3.716$~\cite{leskovec2009community,zhao2014identifying}. We further choose the transmission rate $\lambda$ such that the final epidemic size is that of the well-mixed population without vaccination. For this purpose, we let $\lambda\approx0.085$ for the email network. According to the Refs.~\cite{cornforth2011erratic,luce2008cost,galvani2007long}, the cost of vaccination is set $c_V=\$27$~\cite{luce2008cost} and the cost of infection is set $c_I=\$73$~\cite{galvani2007long}.

\begin{figure}
\begin{center}
\includegraphics[width=3.5in]{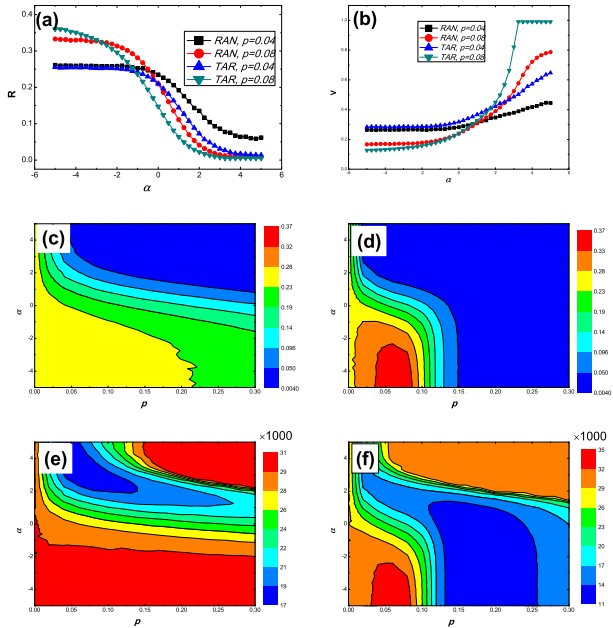}
\caption{(Color online) Comparison between the random subsidy policy and the targeted subsidy policy by considering an influenza case on an email network. (a) the final epidemic size $R$ as a function of $\alpha$; (b) the vaccination coverage $V$ as a function of $\alpha$; (c) the final epidemic size $R$ as a function of ($\alpha,p$) for the random case; (d) the final epidemic size $R$ as a function of ($\alpha,p$) for the targeted case; (e) the social cost $SC$ as a function of ($\alpha,p$) for the random case; (f) the social cost $SC$ as a function of ($\alpha,p$) for the targeted case. The detailed definitions and the values of the other parameters are given in the main text.}
\label{fig7}
\end{center}
\end{figure}

With the above preparations, we compare the two subsidy policies with regard to the vaccination coverage, the final epidemic size as well as the social cost in Fig.~\ref{fig7}. As shown in Fig.~\ref{fig7}, all the phenomena discussed in main text can be observed. Such as, for the targeted policy, decreasing the value of $\alpha$ leads to the rapid reduction of vaccination coverage and yields worse control outcomes (see Fig.~\ref{fig7}(a) and (b)), increasing the value of $p$ can increase the final epidemic size (see Fig.~\ref{fig7}(d)) as $\alpha<0$; For the random subsidy policy, the optimal value of $p$ leading to the minimal social cost appears only when $\alpha>0$, but for the targeted case, the optimal phenomenon can appear for $\alpha>0$ as well as $\alpha\leq0$. Furthermore, one case is worthy to be stressed: for the targeted case, there exists a worst region in which the social cost is the maximal as $p\in(0.03,0.08)$ and $\alpha\in[-5,-2]$.

\section{Conclusions} \label{sec:conclusion}

Targeted immunization strategies have proven to be an efficient measure to control the spread of epidemics on complex networks with heterogenous degree distribution~\cite{pastor2002immunization,zhang2010hub,zhang2014suppression}. However, the strategy assumes that the vaccination campaign is compulsively enforced and the public resources required for vaccination are sufficient. A more realistic case is that the public resources are limited and the government or health organizations can only subsidize a small proportion of individuals --- with the remaining individuals left to decide to vaccinate or not based on their self-interest~\cite{bhattacharyya2012mathematical}. In this situation, we want to know whether the targeted subsidy policy is still better than the random subsidy policy. In this paper, by introducing a preferential selection rule into the epidemiological model, and via a game-theoretic framework, we have studied how the preferential selection rule affects the impact of the targeted subsidy policy and the random subsidy policy on the vaccination coverage, the final epidemic size as well as the social cost. Our findings have shown that, when the selection parameter $\alpha>0$, the targeted subsidy policy can produce positive effects on controlling the spreading of epidemic since these subsidized hubs' vaccination strategy is more likely to be imitated by others. Otherwise, owing to the facts that, on the one hand, the targeted subsidy policy \emph{indirectly} reduces the risk of the remaining nodes and yields the lower vaccination willingness of these nodes, on the other hand, the non-subsidized nodes who are unwilling to take vaccination are more likely to be imitated when the selection parameter $\alpha\leq0$, the combing effects cause the control effect of the targeted policy is worse than the random case. In particular, the introduction of the targeted policy may increase but not reduce the final epidemic size as $\alpha<0$. When considering the impacts of the two subsidy policies on social cost, we found that, for the random subsidy policy, there exists an optimal value of subsidy proportion $p$ leading to the minimal social cost when the selection parameter $\alpha<0$. However, for the targeted case, there always exists an optimal value of $p$ inducing the minimal social cost irrespective of whether $\alpha\leq0$ or $\alpha>0$, and the optimal value of $p$ becomes larger when the value of $\alpha$ is gradually decreased. To illustrate this model we provide an example based on realistic spread of influenza. The results imply that, on the one hand, how to distribute limited public resource to yield the best outcomes sensitively depends on human behavioral responses; whilst on the other hand,  more subsidized individuals do not necessarily guarantee better control effect.

\section*{Acknowledgments}

This work is funded the National Natural Science Foundation of China
(Grant Nos. 61473001, 91324002, 11331009) and HZ is partially supported by the Doctoral Research
Foundation of Anhui University (Grant No. 02303319). MS is funded by an Australian Research Council Future Fellowship (FT110100896).

%

\end{document}